\documentclass[useAMS,usenatbib]{mn2e}
\usepackage{hyperref}
\usepackage{epsfig}
\usepackage{graphicx}
\usepackage{psfrag}
\usepackage{url}
\usepackage{mathrsfs}

\usepackage{amsmath,amssymb}

\def\eprinttmp@#1arXiv:#2 [#3]#4@{
\ifthenelse{\equal{#3}{x}}{\href{http://arxiv.org/abs/#1}{#1}
}{\href{http://arxiv.org/abs/#2}{arXiv:#2} [#3]}}

\providecommand{\eprint}[1]{\eprinttmp@#1arXiv: [x]@}
\newcommand{\adsurl}[1]{\href{#1}{ADS}}

\title[The Cosmological Impact of Intrinsic Alignment Model Choice]
{The Cosmological Impact of Intrinsic Alignment Model Choice for Cosmic Shear}
\author[Donnacha Kirk, Anais Rassat, Ole Host, Sarah Bridle]{Donnacha Kirk$^{1,2}$, Anais Rassat$^{3,4}$, Ole Host$^{1}$, Sarah Bridle$^{1}$\\
$^{1}$Department of Physics \& Astronomy, University College London, Gower Street, London, WC1E 6BT, UK\\
$^{2}$Astrophysics Group, Blackett Laboratory, Imperial College, Prince Consort Road, London, SW7 2BZ, UK\\
$^{3}$LASTRO, Swiss Federal Institute of Technology in Lausanne (EPFL), Observatoire de Sauverny, CH-1290, Versoix, Switzerland\\
$^{4}$Laboratoire AIM, UMR CEA-CNRS-Paris 7, Irfu, SEDI-SAP, Service d'Astrophysique, CEA Saclay, F-91191 Gif s/ Yvette, France}

\begin {document}

\pagerange{\pageref{firstpage}--\pageref{lastpage}} \pubyear{2009}

\maketitle

\label{firstpage}

\begin{abstract}
We consider the effect of galaxy intrinsic alignments (IAs) on dark energy constraints from weak gravitational lensing.
We summarise the latest version of the linear alignment model of IAs, following the brief note of Hirata \& Seljak (2010) and further interpretation in Laszlo et al. (2011).
We show the cosmological bias on the dark energy equation of state parameters $w_0$
and $w_a$ that would occur if IAs were ignored.
We find that $w_0$ and $w_a$ are both catastrophically biased, by an absolute value of just greater than unity under the Fisher matrix approximation.
This contrasts with a bias several times larger for the earlier IA implementation. 
Therefore there is no doubt that IAs must be taken into account for future Stage III experiments and beyond.
We use a flexible grid of IA and galaxy bias parameters as used in previous work, and investigate what would happen if the universe used the latest IA model, but we assumed the earlier version. We find that despite the large difference between the two IA models, the grid flexibility is sufficient to remove cosmological bias and recover the correct dark energy equation of state. In an appendix, we compare observed shear power spectra to those from a popular previous implementation and explain the differences.
\end{abstract}


\section{Introduction}

The gravitational lensing of distant galaxy images has the potential to be a powerful cosmological tool. The lensing effect directly probes the matter distribution as a function of redshift, and thus tells us 
about the expansion history
and growth of structure 
in the Universe.
In this way it allows us to constrain the dark energy equation of state, or whatever is causing the apparent accelerated expansion. 

Weak gravitational lensing poses a number of tough technical challenges if its true potential is to be exploited. The typical cosmic shear induced on galaxies of interest 
 is of order 1\%, which is significantly smaller than the intrinsic ellipticity of the galaxies themselves. Of course we, as observers, have no access to the \textit{unlensed} galaxy images so we must treat a population of galaxies statistically to recover the cosmological information contained in the cosmic shear signal.
The correlation function of galaxy shapes as a measure of gravitational lensing was first proposed by~\citet{kaiser92} and first observed by \citet{Bacon:2000yp,Kaiser:2000if,Wittman:2000tc} and \citet{van_Waerbeke:2000rm}.  
For reviews see \citet{Bartelmann:1999yn,Munshi:2006fn,Refregier:2003ct},
and \citet{hoekstra_jain_2008}.  

A naive approach to cosmic shear assumes that the intrinsic distribution of galaxy ellipticities is random across the sky. If this was 
 the case observed ellipticities on a certain patch of sky could be averaged to recover the cosmic shear, because the intrinsic ellipticity would average to zero. However it was soon pointed out that this assumption of random intrinsic ellipticity distribution is
unjustified 
\citep{HRH,catelankb01,crittendennpt01,croftm00}. 

In fact galaxies
may be expected to align with the large scale gravitational potentials in which they form so we expect physically close galaxies to be preferentially aligned with each other (known as the Intrinsic-Intrinsic (II) correlation). \citet{hiratas04} noted an additional negative correlation between foreground galaxies shaped by a particular gravitational potential and background galaxies which are lensed by the same potential (known as the Gravitational-Intrinsic (GI) correlation) which
can be 
of greater magnitude than the II term.

After the alignment of galaxy ellipticities from linear response to the gravitational potential 
 was proposed as an effect 
  by \citet{catelankb01} the study was put on a firm analytic footing by the introduction of the Linear Alignment (LA) model by \citet{hiratas04}, hereafter HS04. This approach, in which the orientation of galaxies responds linearly to the large-scale gravitational potential in which they form, has become the standard model when including the effects of Instrinsic Alignments (IA).

Correlated shears of galaxies have been observed at low redshift, where they can be attributed to intrinsic alignments, by Brown et al. 2002, and have been measured in galaxies selected to be physically close in \citet{mandelbaumea06,hirataea07,Okumura:2008bm,brainerdea_2009} and \citet{mandelbaumea06}. 
Number density - shear correlations are easier to observe since the random galaxy ellipticity only enters the calculation once, and this has been constrained in \citet{mandelbaumea06,hirataea07,Okumura:2008,mandelbaumea09} and \citet{joachimiea_megazlrg}.

\cite{hirataea07} noted that the scale dependence of the signal they measured was better matched to theory if the non-linear matter power spectrum was used in the LA instead of the linear power spectrum implied in HS04.
\cite{bridleandking} then used the non-linear matter power spectrum in their cosmological forecast calculations.
This approach has been called the Non-Linear Alignment (NLA) ansatz. \citet{schneiderb09} attempted a more motivated solution for assigning power to the IA model at small scales through the use of the halo model of galaxy clustering. By design their halo model reproduced the LA model at large scales.

The II contribution to intrinsic alignments can be taken into account in cosmological analyses by removing galaxies with small physical separation from the analysis e.g. \citet{kings02,kings03}; \citet{heymansh03}; \citet{takadaw04}. This was attempted in a real analysis of COSMOS data in \citet{schrabbackea09} by removing the autocorrelation tomographic bin.
An extension of this method for GI was suggested in  \citet{king05} and developed to a sophisticated level in \citet{joachimi_schneider_2008,joachimi_schneider_2009}. Alternatively a model may be assumed for all the intrinsic alignment contributions, and the free parameters can be marginalised over, as demonstrated at the Fisher matrix level in
\citet{bridleandking,bernstein_2008,joachimi_bridle_2009,MGpaper1,MGpaper2} and demonstrated on real data in \citet{kirk_bs_2010}. 

The redshift evolution of the IA contributions in the LA model 
was found to be incorrect 
due to a mistake in HS04 in the conversion between the primordial potential and the matter power spectrum.
This
was corrected in a new version of HS04, issued as \citet{hiratas10_posterratum},
hereafter HS10. 
In addition there is an ambiguity in HS04 as to which cosmological epoch is responsible for the ``imprinting'' of galaxy intrinsic alignments. Most starkly the question is, are IAs frozen in at some redshift of formation or do they evolve with the growth of structure, particularly nonlinear clustering on small scales? \citet{blazek2011} note this as an issue in the HS04 approach and estimate its effect on the GI amplitude as of order 20\%. \citet{MGpaper1}, \citet{MGpaper2} implement one physically motivated solution to this question (as well as including the redshift evolution correction of \citealt{hiratas10_posterratum}) which we adopt in this work.

The result of this evolution in the treatment of IAs is that much useful theoretical
and observational work has been conducted using an incorrect implementation of the LA model,
using often unjustified treatments of the small scale
seeding 
of galaxy IAs. In this work our aim is to present basic results for the most up to date implementation of the LA model for IAs. We present angular power spectra for components of the shear-shear ($\epsilon\epsilon$), position-position ($nn$) and position-shear ($n\epsilon$) observables as well as the reduction in constraining power in measuring dark energy caused by a robust treatment of IAs and the biasing of cosmological parameters that results from ignoring IAs or treating them using an old model. Many of these results reproduce previous work in the literature which was conducted using an old implementation of the LA model. Specifically we reproduce Fig. 1 from \citet{joachimi_bridle_2009} for the new implementation of the LA model which we hope will act as a reference for those wishing to apply it in the future.
Furthermore we investigate the ability of a flexible parameterisation of intrinsic alignments to compensate for using the old LA model compared to the latest version we summarise here.

This paper is organised as follows: In section \ref{sec:LA_now} we describe the most up to date implementation of the LA model of IAs, detailing our formalism for shear-shear ($\epsilon\epsilon$), position-position ($nn$) and position-shear ($n\epsilon$) correlations.
Section \ref{sec:bias} presents the bias on cosmological parameter estimation caused by mistreatment of IAs for the latest model, the old model and an intermediate model; we then conclude in section \ref{sec:conclusions}.
In an appendix we summarise the history of the LA model and present the major differences between the latest implementation and the previous widely used version.

\section{The Linear Alignment Model}
\label{sec:LA_now}

In this section we summarise the latest linear alignment model interpretation and its contribution to (i) shear-shear correlations (ii) position-position correlation and (iii) position-shear cross correlation.

\subsection{Shear-shear correlations, $\epsilon\epsilon$}
\label{sec:LA_ee}

The Linear Alignment (LA) model for galaxy intrinsic ellipticities was suggested by \citet{catelankb01} and applied quantitatively by \citet{hiratas04}, see section \ref{sec:LA_history} below for a more complete history. It assumes that, under gravitational collapse, dark matter forms into triaxial halos whose major axis is aligned with the
maximum curvature of the 
large scale gravitational potential.
Elliptical galaxies are expected to trace the ellipticity of their parent halos, meaning that the ellipticity distribution of a population of elliptical galaxies will be linearly related to the
curvature of the 
gravitational potential, $\phi$, via
\begin{eqnarray}
\epsilon_{+} &=& C (\partial_{y}^{2} - \partial_{x}^{2}) \phi \\
\epsilon_{\times} &=& 2C\partial_{y}\partial_{x} \phi,
\end{eqnarray}
where $C$ is
an unknown constant, fixed for our formalism by a normalisation given below.

In Weak Gravitational Lensing (WGL) we are interested in IAs as systematic effects which contaminate a measured cosmic shear signal, specifically the shear-shear power spectrum. The LA model motivates the amplitude of two separate
terms
which affect our ability to accurately measure cosmic shear. The first, Intrinsic-Intrinsic (II), term arises when galaxies are physically close. These galaxies form in the same gravitational potential, hence their intrinsic ellipticities are preferentially aligned. This produces a spurious positive correlation which adds to the measured cosmic shear signal.

The second,
Gravitational-Intrinsic (GI),
term becomes important when treating galaxies which are close on the sky but separated in redshift. In this case foreground galaxies will align to a foreground gravitational potential which will itself contribute to the gravitational lensing of the background galaxies. This produces an anti-correlation in observed ellipticities and subtracts from the
expected cosmic shear signal.

In terms of projected angular power spectra we can write the measured ellipticity power spectrum, $C_{l}^{\epsilon\epsilon}$, as a sum of the
gravitational lensing 
shear power spectrum and two IA terms
\begin{equation}
C^{\epsilon\epsilon}(l) = C^{GG}(l) + C^{II}(l) + C^{GI}(l). 
\end{equation}

The cosmic shear power spectrum, $C_{ij}^{GG}(l)$, under the Limber approximation, is written as the integrated product of the matter power spectrum, $P_{\delta\delta}(k,\chi)$, and the lensing weight function, $W_{i}(\chi)$
\citep{hu99}, 
\begin{equation}
C_{ij}^{GG}(l) = \int_{0}^{\chi_{\textrm{hor}}} \frac{\textrm{d}\chi}{\chi^2}W_{i}(\chi)W_{j}(\chi)P_{\delta\delta}(k,\chi),
\end{equation}
where $P_{\delta\delta}(k,\chi) \equiv P_{\delta\delta}(\frac{l}{f_{K}(\chi)},\chi)$, $\chi$ is the comoving distance along the line of sight,$f_{K}(\chi)$ is the comoving angular diameter distance, $i,j$ denote a tomographic redshift bin pair and
\begin{equation}
W_{i}(\chi) = \frac{3}{2}\frac{H_{0}^{2}\Omega_m}{c^2}\frac{\chi}{a}\int_{\chi}^{\chi_{\mathrm{hor}}}\textrm{d}\chi'n_{i}(\chi')\frac{\chi' - \chi}{\chi'},
\end{equation}
with $H_0$ the Hubble parameter today, $\Omega_m$ the dimensionless matter energy density, $c$ the speed of light, $a$ the dimensionless scale factor and $n_{i}(\chi')$ is the galaxy redshift distribution for a particular bin $i$.

Similarly, the 
projected angular power spectra for the IA terms are \citep{hiratas04}
\begin{eqnarray}
C_{ij}^{II}(l) &=& \int_{0}^{\chi_{\textrm{hor}}} \frac{\textrm{d}\chi}{\chi^2}n_{i}(\chi)n_{j}(\chi)P_{II}(k,\chi) \\
C_{ij}^{GI}(l) &=& \int_{0}^{\chi_{\textrm{hor}}} \frac{\textrm{d}\chi}{\chi^2}W_{i}(\chi)n_{j}(\chi)P_{GI}(k,\chi),
\label{eqn:Cl_II_GI_new}
\end{eqnarray}
where, unsurprisingly, the II correlation depends only on the galaxy redshift distribution, $n(\chi)$, while GI depends on the product of the galaxy redshift distribution and the lensing weight function.

The IA ``power spectra'', $P_{II}(k,\chi)$ and $P_{GI}(k,\chi)$,
are unknown functions of scale and cosmic epoch. 
In the case of the LA model they become
\begin{eqnarray}
P_{II}(k,\chi) &=& \left( -C_1 \rho(\chi=0) \right)^2 P_{\delta\delta}^{\rm lin}(k,\chi=0) \label{eqn:P_II_GI_new1} \\
P_{GI}(k,\chi) &=& -C_1 \rho(\chi=0) \sqrt{P_{\delta\delta}^{\rm lin}(k,\chi=0)} \frac{\sqrt{P_{\delta\delta}(k,\chi)}}{D(z)} \label{eq:sqrtpk}
\label{eqn:P_II_GI_new2}
\end{eqnarray}
where $P_{\delta\delta}$ is the matter power spectrum, $\rho(\chi=0)$ is the matter density today and $D(z)$ is the linear growth factor
as a function of redshift $z$, where it is implicitly meant that $z=z(\chi)$. 
$C_1$ is the normalisation of the IA contribution and
we use a fiducial value of 
$5\times 10^{-14}(h^2 M_{\odot}Mpc^{-3})^{-1}$, following \citet{bridleandking} who match to the power spectra in
\citet{hiratas04} which are based on SuperCOSMOS data from \citet{brownthd02}.

The II term is related to the linear matter power spectrum, $P_{\delta\delta}^{\rm lin}$, because we assume that galaxy intrinsic ellipticity is imprinted at the (early) epoch of galaxy formation and subsequently ``frozen in''. As such the power spectrum of its distribution does not undergo the nonlinear evolution that the distribution of matter clustering does at late times. The factor of $\sqrt{P_{\delta\delta}^{\rm lin}(k,\chi=0)P_{\delta\delta}(k,\chi)}/D(z)$ in the GI term reflects the dependence on both the intrinsic ellipticity distribution and the full, late time, mass distribution via gravitational lensing. The exact relation between intrinsic ellipticity, the epoch of galaxy formation and the evolution of galaxy clustering is a vexed and disputed one. A more detailed discussion of the history of the treatment of this relationship can be found in the appendix below. We believe that our approach best reflects the physical understanding of the LA model as proposed by \citet{hiratas04} who related the intrinsic ellipticity distribution to the Newtonian potential at the time of galaxy formation.

\subsection{Position-position correlations, nn}
\label{sec:LA_nn}

Any cosmic shear survey provides not only a catalogue of measured galaxy ellipticities but also 
 a record of galaxy positions through their location on the sky and an estimate of their redshift. Analogously to the cosmic shear power spectrum we may define the galaxy position
density 
power spectrum,
$C^{nn}(l)$,
the Fourier transform of the galaxy position two-point correlation function. It is conceptually useful to divide the contributions up as follows,
\begin{equation}
C^{nn}(l) = C^{gg}(l) + C^{mm}(l) + C^{gm}(l).
\end{equation}
Here we have separated the contribution from galaxy clustering itself, $gg$, from the contribution due to lensing magnification, $mm$, and the cross-correlation of the two, $gm$. The lensing contribution manifests because a galaxy
may increase (or decrease) in size as a result of the image distortion. The application of Liouville's theorem tells us that the process of WGL conserves surface brightness density hence a larger (smaller) galaxy will appear brighter (dimmer) due to the lensing effect. For a magnitude limited survey this can affect the statistics of galaxy clustering as otherwise too faint galaxies are 
 promoted into the survey by magnification and vice-versa. As the $nn$ correlation is effectively just a counting of galaxies and their relative positions there is no ellipticity contribution and we can ignore IAs.

The equation for the pure galaxy clustering term is relatively straightforward
in the Limber approximation 
\begin{equation}
C_{ij}^{gg}(l) = \int_{0}^{\chi_{\textrm{hor}}} \frac{\textrm{d}\chi}{\chi^2}n_{i}(\chi)n_{j}(\chi) b_{g}^{2}(k,z)P_{\delta\delta}(k,\chi),
\end{equation}
where the window function is the galaxy redshift distribution as we would expect and we have introduced the galaxy bias term, $b_{g}$, to reflect our understanding that galaxies are a biased tracer of the underlying matter distribution. As before, $P_{\delta\delta}(k,\chi) \equiv P_{\delta\delta}(\frac{l}{f_{K}(\chi)},\chi)$.
In general $b_{g}$ is an unknown function of scale and cosmic epoch. 
The details of
the 
galaxy bias formalism
we assume in this paper, 
along with other bias terms, are explained in section \ref{sec:bias} below.

The magnification power spectrum and galaxy clustering-magnification cross-term can be written
\begin{eqnarray}
C_{ij}^{mm}(l) &=& 4(\alpha_{i}-1)(\alpha_{j}-1)C_{ij}^{GG}(l) \\
C_{ij}^{gm}(l) &=& 2(\alpha_{j}-1)C_{ij}^{gG}(l),
\end{eqnarray}
where $\alpha_{i}$ is defined as the slope of the luminosity function, evaluated at the median redshift of the bin $i$, and $C^{gG}(l)$ is defined in section \ref{sec:LA_ne}. In this approach we are following \citet{joachimi_bridle_2009}.
We treat 
each of the $\alpha_{i}$ terms as a free parameter in the model,
and take fiducial parameters from eqns. 32, 33 and Table 2 of \citet{joachimi_bridle_2009}
(see Appendix A of that paper for more details).

\subsection{Position-shear correlations, $n\epsilon$}
\label{sec:LA_ne}

As well as $\epsilon\epsilon$ and $nn$ correlations themselves,
we can cross-correlate the two fields to form the  
galaxy position-shear, $n\epsilon$, cross-correlation
functions as first proposed by \citet{huj04} in the context of dark energy. 
This is often referred to as galaxy-galaxy lensing, in which the mass of a foreground galaxy distorts the shape of a background galaxy. 
Here we consider the general cross-correlation which includes contributions from larger dark matter structures, from magnification and also from intrinsic alignments if the correlated galaxies are physically close. 
We can write the contributions to the full $n\epsilon$ term as
\begin{equation}
C^{n\epsilon}(l) = C^{gG}(l) + C^{gI}(l) + C^{mG}(l) + C^{mI}(l)
\end{equation}
\citep{joachimi_bridle_2009}.

The individual expressions for the terms in the
$C^{n\epsilon}(l)$ 
expansion
contain combinations of quantities already considered 
\begin{eqnarray}
C_{ij}^{gG}(l) &=& \int_{0}^{\chi_{\textrm{hor}}} \frac{\textrm{d}\chi}{\chi^2}n_{i}(\chi)W_{j}(\chi)\nonumber\\
&& b_{g}(k,z)r_{g}(k,z)P_{\delta\delta}(k,\chi) \\
C_{ij}^{gI}(l) &= & \int_{0}^{\chi_{\textrm{hor}}} \frac{\textrm{d}\chi}{\chi^2}n_{i}(\chi)n_{j}(\chi) \nonumber\\
&& b_{g}(k,z)r_{g}(k,z)b_{I}(k,z)r_{I}(k,z)P_{GI}(k,\chi) \\
C_{ij}^{mG}(l) &= & 2(\alpha_{i}-1)C_{ij}^{GG}(l) \\
C_{ij}^{mI}(l) &= & 2(\alpha_{i}-1)C_{ij}^{GI}(l),
\end{eqnarray}
where we have introduced the additional free functions $r_{g}$ and $r_{I}$ which appear 
 due to the possible stochastic relation between the galaxy and dark matter distributions \citep{dekell99} as discussed in section \ref{sec:bias} below. As before $P_{\delta\delta}(k,\chi) \equiv P_{\delta\delta}(\frac{l}{f_{K}(\chi)},\chi)$.
We note that $P_{GI}(k,\chi)$ appears in the $C_{ij}^{gI}(l)$ term because it describes how the dark matter distribution relates to the intrinsic alignments, and we assume the relation between the galaxy distribution and the dark matter distribution can be accounted for by the product $b_{g} r_{g}$.

\subsection{Summary of Observables and Fields}

\begin{figure*}
  \begin{flushleft}
    \centering
       \includegraphics[width=7in,height=7in]{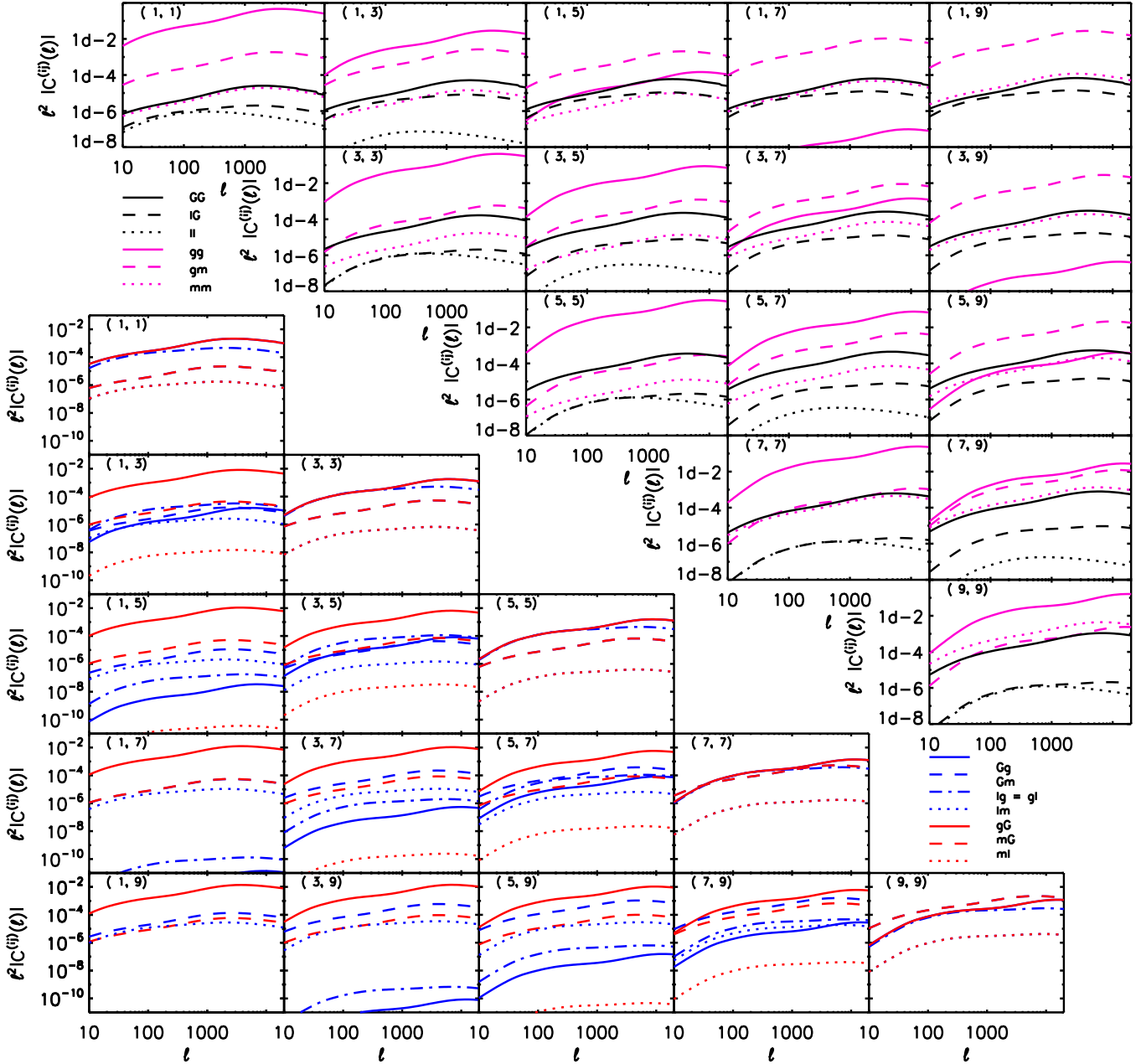}
\vspace{6mm}
\caption{Fiducial power spectra for all correlations considered. The upper right panels depict the contributions to the $\epsilon\epsilon$ (in black) and nn (in magenta) correlations. The lower left panels show the contributions to correlations between number density fluctuations and ellipticity. Since we only show correlations $C_{\alpha\beta}^{ij}(l)$ with $i \leq j$, we make in this plot a distinction between $n\epsilon$ (in red; number density contribution in the foreground, e.g. gG) and $\epsilon n$ (in blue; number density contribution in the background, e.g. Gg) correlations. In each sub-panel a different tomographic redshift bin correlation is shown. For concision only odd bins are displayed. In the upper right panels the usual cosmic shear signal (GG) is shown as black solid lines; the intrinsic alignment GI term is shown by the black dashed lines; the intrinsic alignment II term is shown by the dotted black lines; the usual galaxy clustering signal (gg) is shown by the magenta solid lines; the cross correlation between galaxy clustering and lensing magnification (gm) is shown by the magenta dashed lines; the lensing magnification correlation functions (mm) are shown by the magenta dotted lines. In the lower left panels the
solid blue lines show the correlation between lensing shear and galaxy clustering (Gg); the blue dashed lines show the correlation between lensing shear and lensing magnification (gm); the blue dot-dashed lines show the correlation between intrinsic alignment and galaxy clustering (Ig or equivalently gI); the red solid lines show the correlation between galaxy clustering and lensing shear (gG), which is equivalent to the blue solid lines with redshift bin indices i and j reversed; similarly the red dashed lines show the correlation between lensing magnification and lensing shear (mG), for cases where the magnification occurs at lower redshift than the shear ($i < j$); finally the dotted lines show the correlation between lensing magnification and intrinsic alignment (mI).}
\label{fig:cls_full}
  \end{flushleft}
\end{figure*}

The observables considered in this paper and the different fields which contribute to each observable are summarised in Table \ref{tab:summary}. In Fig.~\ref{fig:cls_full} we plot the angular power spectra of all considered components for a fiducial stage-IV lensing survey. For our fiducial model 
in this paper we use the following survey specification: 
20,000 deg$^2$ with a galaxy number density of 35 arcmin$^{-2}$ and a redshift distribution given by the Smail-type
$n(z)$,
\begin{equation}
n(z)=z^{\alpha}\exp{\left(- \left( \frac{z}{z_{0}} \right)^{\beta}\right)},
\end{equation}
with $\alpha=2$, $\beta=1.5$, $z_0=0.9/\sqrt{2}$, divided into 10 tomographic bins with equal number density out to redshift 3 and normalised so that $\int n(z) dz = 1$. A Gaussian photometric redshift error of $\sigma_z = 0.05(1+z)$ is assumed with no catastrophic outliers in redshift. We constrain the set of cosmological parameters $\left\{ \Omega_m, w_0, w_a, h, \sigma_8, \Omega_b, n_s \right\}$ which take the values $\Omega_m=0.25$, $w_0=-1$, $w_a=0$, $h=0.7$, $\sigma_8=0.8$, $\Omega_b=0.05$, $n_s=1$. All results assume a flat, $\Lambda$CDM cosmology. Linear matter power spectrum is calculated using the \citet{eisensteinhu97} fitting function and nonlinear corrections are applied  where appropriate using the \citet{smithea03} formalism. 

\begin{table}
   \centering
   \begin{tabular}{@{} l|l @{}} 
   \hline
Observables\\

$\epsilon \epsilon$& shear-shear\\
$nn$ & position-position\\
$n\epsilon$ & position-shear\\
\hline
Fields\\

$G$ & gravitational lensing\\
$I$ & intrinsic alignment\\
$g$ & galaxy clustering\\
$m$ & cosmic magnification\\
\hline
   \end{tabular}
   \caption{Summary of observables considered in this paper and different fields which contribute to each observable.}
   \label{tab:summary}
\end{table}

Fig.~\ref{fig:cls_full} is presented and formatted in the same way as Fig. 3 in \citet{joachimi_bridle_2009}. The purpose is to supply a new reference plot using the most up to date implementation of the LA model for IAs. All lines which do not include an ``I'' term will be identical in both versions of the plot. However, as \citet{joachimi_bridle_2009} use the original HS04 implementation, from before the publication of the erratum incorporated into \citet{hiratas10_posterratum}, and present lines for the NLA ansatz, there
are differences in any lines containing an I contribution. The NLA version of HS04 is the most widely used in the literature to date; we refer to it subsequently as HS04NL.
Note that 
that we are employing the NLA prescription in HS04NL even though this was not explicitly discussed in the original \citet{hiratas04} paper. See the appendix for a more detailed explanation of the evolution of the LA model and the differences between various implementations.

The upper triangle of Fig. \ref{fig:cls_full} shows the contributions to the shear-shear correlation function in black for every other tomographic bin pairing. As usual the lensing contribution is the largest on all angular scales, 
but is most dominant in the auto-correlations at high redshift. The II intrinsic alignment term is most important in the auto-correlations and negligible for widely spaced bins. The GI contribution is largest for separated bins (shown as IG for compactness - see caption).  The intrinsic alignment contributions are typically one to ten per cent of the lensing signal.

The contributions to the position-position correlation function are shown in pink in the upper triangle and as expected, the auto-correlations are dominated by the intrinsic clustering of galaxies. The cross-correlations of separated bins can be dominated by the cross-correlation of galaxy clustering and magnification. As expected, the magnification-only term is most important at high redshift, although it never exceeds the other terms for the redshift ranges we consider.

The lower triangle shows the various contributions to the position-shear cross-correlation function, and shows the effect of reversing the tomographic bin order where relevant 
(see caption).
The correlation between lensing shear and galaxy clustering is the strongest of these contributions and is largest when the tomographic bins are separated with the galaxies in the foreground. This is the usual galaxy-galaxy lensing contribution. Interestingly it is still large in the auto-correlations which is due to the galaxies in the nearest part of the redshift bin lensing the galaxies in the farthest part. There is a non-negligible contribution in some of the cross terms from the lensing of galaxies in a nearer tomographic bin by galaxies in a farther tomographic bin, due to the overlap in redshift distributions (solid blue lines). At highest redshifts the magnification terms start to dominate. The cross-correlation between magnification and intrinsic alignment is mostly one of the smallest terms.

\section{Biasing of Cosmological Parameters}
\label{sec:bias}

The LA model is currently our best description of galaxy intrinsic alignment over a range of cosmologies. It has been shown that ignoring IAs in an analysis can significantly bias the measurement of fundamental cosmological parameters \citep{bridleandking}. 
 In this section we quantify this bias and examine the effect of moving from the old standard LA implementation to the new one. We also explore the bias produced by using the old implementation rather than the new.

A naive approach to cosmic shear measurements would ignore IAs and measure values for cosmological parameters which are systematically biased. By employing a flexible IA model and marginalising over a set of nuisance parameters we lose precision but hope to produce unbiased cosmological measurements. In this section we quantify the cosmological bias which results from ignoring IAs and also the cosmological bias which results from employing the HS04NL LA model rather than the current LA model.

To this end we employ the cosmological bias formalism of
\citet{huterertbj06} (see also \citet{amarar07} and appendix A of \citet{joachimiea_megazlrg}). 
\begin{equation}
\delta p_{\alpha} = F_{\alpha\beta}^{-1} \sum_{l} \Delta C_{ij} \left( {\rm Cov } \left[ C_{ij}(l),C_{mn}(l) \right] \right)^{-1} \frac{\partial C_{mn}(l)}{\partial p_{\beta}},
\end{equation}
where $\delta p_{\alpha}$ is the cosmological bias on each parameter considered, $C_{ij}$ are the projected angular power spectra, $\frac{\partial C_{mn}(l)}{\partial p_{\beta}}$ are the derivatives of these power spectra with respect to the cosmological parameters, ${\rm Cov} \left[ C_{ij}(l),C_{mn}(l) \right]$ is the covariance matrix of the power spectra, calculated according to eqn. 39 
 in \citet{joachimi_bridle_2009}, and $F_{\alpha\beta}^{-1}$ is the inverse of the Fisher Matrix (FM) for our set of cosmological parameters, calculated as
\begin{equation}
F_{\alpha\beta} = \sum^{l_{max}}_{l=l_{min}} \sum_{(i,j),(m,n)} \frac{\partial C_{ij}(l)}{\partial p_{\alpha} }{\rm Cov}^{-1} \left[ C_{ij}(l),C_{mn}(l) \right] \frac{\partial C_{mn}(l)}{\partial p_{\beta} }.
\nonumber 
\end{equation}
 Each of
the FM terms is calculated using the assumed IA model. $\Delta C_{ij}$ is the difference between the data vector calculated using the assumed model and the true model,
\begin{equation}
\Delta C_{ij} = C^{\textrm{assumed}}_{ij}(l) - C^{\textrm{true}}_{ij}(l).
\end{equation}

\begin{figure}
  \begin{flushleft}
    \centering
       \includegraphics[width=3.5in,height=3.5in]{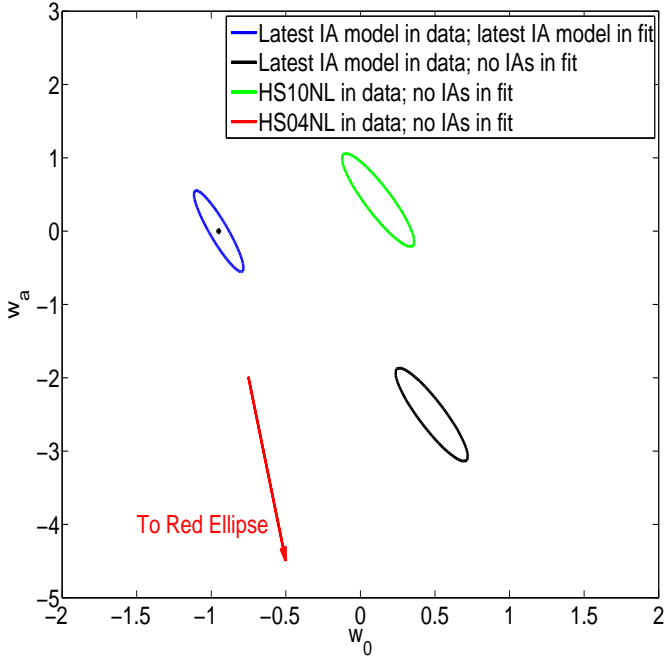}
\caption{95\% confidence limits on the dark energy parameters $w_0$ and $w_a$ for the IA implementation from this work [blue contour] and the biased constraints if it is assumed that IAs do not exist but in fact the new implementation [black contour], the HS10NL implementation [green contour] or the HS04NL implementation [off plot, direction indicated by red arrow] are true. The cosmological bias is given by the offset between each displaced contour and the fiducial values of $(-1,0)$ [black cross]. Results are presented for an observable data vector of shear-shear ($\epsilon\epsilon$) correlations only. It is assumed that we enjoy perfect knowledge of the IA contribution for each implementation.}
\label{fig:bias_ignore}
  \end{flushleft}
\end{figure}

Fig. \ref{fig:bias_ignore} shows the cosmological bias on the dark energy equation of state parameters if IAs are assumed not to exist but they are truly present according to one of: the HS04NL implementation; the new model; or an intermediate between the two, called HS10NL. The bias is the distance between the centre of the ``true'' contour [-1,0] and the centre of the contours which assume no IAs.

\begin{figure}
  \begin{flushleft}
    \centering
       \includegraphics[width=3.5in,height=3.5in]{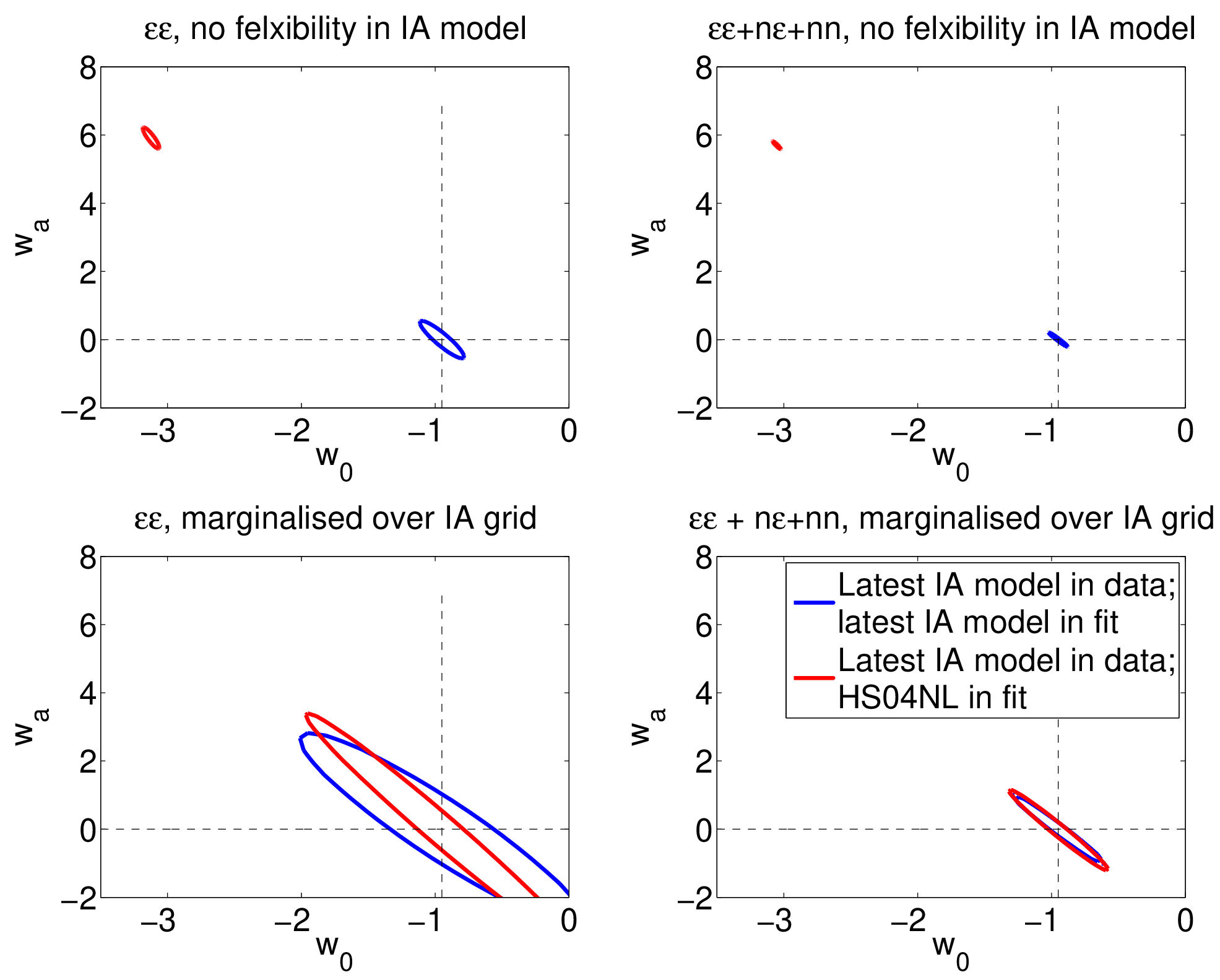}
\caption{95\% confidence limits on the dark energy parameters $w_0$ and $w_a$ for the IA implementation from this work [blue contours] and the biased constraints if the old HS04NL implementation is assumed [red contours]. The cosmological bias is given by the offset between the displaced contour and the fiducial values of $(-1,0)$ [black dotted lines]. Results are presented for an observable data vector of shear-shear ($\epsilon\epsilon$) correlations alone [left-hand panels] and the full combination of shear-shear,shear-position and position-position ($\epsilon\epsilon+n\epsilon+nn$) correlations [right-hand panels]. Each probe combination is shown for the case of perfect knowledge of IAs and galaxy bias [top panels] and for the case where our ignorance is accounted for by marginalisation over the nuisance parameter grid with $N_k=N_z=7$ [bottom panels].}
\label{fig:bias_oldnew}
  \end{flushleft}
\end{figure}

Note that the bias parameterisation used is based on the FM formalism which assumes purely Gaussian errors in the parameters 
 and observables which respond linearly with respect to the parameters in question.
The parameterisation
is very likely to break 
 down for parameter values more than $\sim 2-3 \sigma$ away from the fiducial cosmology. As such the more extreme bias values seen in Figs. \ref{fig:bias_ignore} and \ref{fig:bias_oldnew} should not be read as exact. What they can tell us is rough relative bias between different scenarios and it is clear that any scenario producing an absolute bias on $w_0$ of order unity or above can be considered to be ``catastrophically biased''. For clarity we use the notation $\sigma_{\textrm{FM}}$ when quoting the size of biases on parameters with respect to the errors on the unbiased constraints to remind the reader that they are calculated assuming the Fisher Matrix formalism is valid.  

The intermediate model (HS10NL) is introduced to disentangle the two effects which change between HS04NL and the new LA implementation. It includes the correct factors of $a$, introduced by \citet{hiratas10_posterratum} and shown in equations \ref{eqn:HS04_P_II_GI_erratum}, but it always applies the non-linear matter power spectrum, $P^{lin}_{\delta\delta}(k,z)$, to the IA terms, following the \citet{bridleandking} interpretation of HS04NL. So this HS10NL model applies the correct redshift evolution to IAs but assumes that the effects of nonlinear clustering are always present, even in the II term, by adopting the NLA ansatz.

Ignoring IAs causes strong biasing on the dark energy parameters, no matter which IA model is assumed to be true.
The new implementation
is the least biased at $\sim 8\sigma_{\textrm{FM}}$ away from the true model.
However, ignoring IAs appears to
still bias $w_0$ by
of order $\sim1.5$.
The HS10NL model is more biased at  $\sim20\sigma_{\textrm{FM}}$. The HS04NL implementation produces the strongest cosmological bias, with the contour far outside the plotted area. At a point this far away from the fiducial parameter values the Gaussian assumptions of the Fisher matrix and bias formalisms certainly break down and it would be wrong to put much faith in the exact direction/distance of the predicted bias. What is clear however is that the effect is very strong. This finding is what we would expect given that the move from HS04NL to the new implementation has reduced the impact of IAs, not only through changes to their redshift evolution, but also through the removal of IA power on small scales. 

The general biasing trend is the same when we calculate for a stage-III type survey, like the Dark Energy Survey (DES), with biases of $\sim 4 \sigma_{\textrm{FM}}$ for the latest model,  $\sim 8 \sigma_{\textrm{FM}}$ for the HS10NL implementation and $\sim 30 \sigma_{\textrm{FM}}$ for the HS04NL model. 

So far we have not taken into account any uncertainty in the IA model, but assumed they are zero in the parameter fitting, whichever model we employ to describe them
in the true model. 
In reality, we are aware that our knowledge of IAs from simulations and observations is still developing and relatively uncertain. The case is similar for galaxy bias, introduced in section \ref{sec:LA_nn}, and employed below. To parameterise our ignorance of both effects and their cross-correlations we use a grid of nuisance parameters
\begin{equation}
X = A_{X}Q_{X}(k,z),
\end{equation}
where $A_X$ is a constant amplitude parameter, free to vary about a fiducial value of 1, and $Q_{X}(k,z)$ is a grid of $N_{k} \times N_{z}$ nodes logarithmically spaced in k/z space, each of which is allowed to vary independently around a fiducial value of 1. A final smooth grid is created by spline interpolation over the values of the grid nodes. For more details of this nuisance parameter grid see \citet{joachimi_bridle_2009}, \citet{MGpaper1}, \citet{MGpaper2}. 
This was inspired by the marginalisation of \citet{Bernstein_2008} which led to the grid implementation in \citet{joachimi_bridle_2009}. 

We define four sets of nuisance parameters which each take the form of this grid,
\begin{eqnarray}
b_{I}(k,z) &= A_{b_{I}}Q_{b_{I}}(k.z) \\ 
b_{g}(k,z) &= A_{b_{g}}Q_{b_{g}}(k.z) \\
r_{I}(k,z) &= A_{r_{I}}Q_{r_{I}}(k.z) \\
r_{g}(k,z) &= A_{r_{g}}Q_{r_{g}}(k.z) .
\end{eqnarray}

The ``$b$" terms can be understood as bias terms between the power spectrum of a field X and the matter power spectrum, i.e.:
\begin{equation} \frac{P_{XX}(k,z)}{P_{\delta \delta}(k,z)} = b^2_X(k,z)\end{equation}
and ``$r$" terms are biasing terms which arise in cross-correlations, i.e.:
\begin{equation} \frac{P_{XY}(k,z)}{P_{\delta \delta}(k,z)} = b_X(k,z)r_X(k,z)b_Y(k,z)r_Y(k,z).\end{equation} The ``$b$" and ``$r$" terms are both functions of the clustering and stochastic bias as described in \cite{dekell99}.

These sets of nuisance parameters then appear, in different combinations, in each angular power spectrum integral where galaxy clustering, $g$, or the IA term, $I$, appear:
\begin{eqnarray}
C_{ij}^{II}(l) &=& \int \frac{\textrm{d}\chi}{\chi^2}n_{i}(\chi)n_{j}(\chi)b_{I}^{2}(k,z)P_{II}(k,\chi) \\
C_{ij}^{GI}(l) &=& \int \frac{\textrm{d}\chi}{\chi^2}W_{i}(\chi)n_{j}(\chi)b_{I}(k,z)r_{I}(k,z)P_{GI}(k,\chi) \\
C_{ij}^{gg}(l) &=& \int \frac{\textrm{d}\chi}{\chi^2}n_{i}(\chi)n_{j}(\chi)b_{g}^{2}(k,z)P_{\delta\delta}(k,\chi) \\
C_{ij}^{gG}(l) &=& \int \frac{\textrm{d}\chi}{\chi^2}n_{i}(\chi)W_{j}(\chi)b_{g}(k,z)r_{g}(k,z)P_{\delta G}(k,\chi) \\
C_{ij}^{gI}(l) &=& \int \frac{\textrm{d}\chi}{\chi^2}n_{i}(\chi)n_{j}(\chi)b_{I}(k,z)r_{I}(k,z) \nonumber\\
&&b_{g}(k,z)r_{g}(k,z)P_{\delta I}(k,\chi),
\end{eqnarray}
while nuisance parameters appear in $C_{ij}^{gm}(l)$ and $C_{ij}^{mI}(l)$ due to their dependence on $C_{ij}^{gG}(l)$ and $C_{ij}^{GI}(l)$ respectively. In general this means that, for the full suite of observables $\epsilon\epsilon + n\epsilon + nn$, we marginalise over a total of $4(1+N_{k}\times N_{z})$ nuisance parameters, 
plus 30 that parameterise the magnification effects. As before $P_{\delta\delta}(k,\chi) \equiv P_{\delta\delta}(\frac{l}{f_{K}(\chi)},\chi)$. 
 When we employ the nuisance grid in this work we set $N_k = N_z = 7$, giving a total of 230 nuisance parameters. This follows the practice used in \citet{joachimi_bridle_2009} for the purposes of comparison (see their Table 2). We also restrict the information used in the galaxy field to linear scales as in \citet{rassatea_08}. 

 This current paper presents the most advanced interpretation of the LA model. However, much work of the work on IAs in the literature has been conducted using
  a non-linear version of 
  the original interpretation presented in \citet{hiratas04}. In Fig. \ref{fig:bias_oldnew} we quantify the bias introduced on the dark energy parameters from the use of this HS04NL approach rather than the latest implementation introduced in this work. 
   When the IA signals (whether from the old or new models) are assumed to be known perfectly there is a very strong biasing effect which would result in a very worrying systematic mis-measurement of cosmology.

 In the case where $\epsilon\epsilon$ correlations alone are considered, the bias is $\sim 25\sigma_{\textrm{FM}}$ of the constraining power of the cosmic shear signal given the true IA contribution. When the full set of correlations including position information, $\epsilon\epsilon+n\epsilon+nn$, are considered this biasing increases to $\sim65\sigma_{\textrm{FM}}$. As previously discussed, when the effect is this strong the exact position/direction of the biasing
is likely to be well 
outside the competence of the FM-based formalism but the fact that there is a very significant effect should not be doubted. 

 When we employ the robust parameterisation of both IA and galaxy bias uncertainties via the grid of nuisance parameters described above we see a striking change in the results. As we would expect the constraining power of either probe combination decreases as we have marginalised over 130 free parameters which represent our ignorance of the exact details of IAs (and another 100 for galaxy bias). What is also clear is that the biasing of our cosmological estimates has decreased to well within the $1\sigma$ error of the respective probes.

 The constraints on $w_0$ and $w_a$ from the full  $\epsilon\epsilon+n\epsilon+nn$ combination after marginalisation over 230 nuisance parameters are only reduced by a factor of two compared to the naive (and heavily biased if the wrong IA model is used) constraints from $\epsilon\epsilon$ alone when it is assumed IAs are perfectly known. This corresponds to a factor of 2.6 reduction in DETF FoM \citep{detf}.

\section{Conclusions}
\label{sec:conclusions}

Galaxy IAs are the most important physical systematic in the study of cosmic shear. As increasing volumes of data become available from WGL surveys more interest is being paid to their correct treatment. 
Much of this work has been dominated by the LA model, originally introduced in HS04 (and the NLA extension
which we refer to as HS04NL). 
 This original implementation was subsequently corrected in \citet{hiratas10_posterratum} and detailed attention paid to the treatment of non-linear clustering in \citet{MGpaper1,MGpaper2}.

However, much of the existing literature has been produced using the uncorrected HS04NL model. In this paper we have provided a brief explanation of the evolution of, and context surrounding, the LA model for IAs, highlighting the most important differences between HS04NL and the latest implementation.

 We have calculated the angular power spectra of the cosmic shear observables, correcting
the implicit mistake of \citet{joachimi_bridle_2009}, which was based on HS04, which we hope will act as a new reference for those interested in applying IAs to their cosmic shear analysis.

The main motivation for the study of IAs is the measurement of unbiased constraints of key cosmological parameters. We show that the new LA implementation significantly reduces the impact of IAs, and hence the bias, but that the effect is still very significant, producing a bias at the tens of $\sigma$ level when we assume perfect knowledge of IAs.

If we did know the IA signal perfectly then we could produce unbiased measurements by simply subtracting the IA signal from our measured cosmic shear. In practice it is useful to parameterise our ignorance of the true IA signal through a set of nuisance parameters which are marginalised over to produce weaker but hopefully unbiased cosmological estimates. We show that a robust grid of 130 nuisance parameters for IAs and magnification uncertainties, 
 allowed to vary in scale and redshift, effectively removes the bias due to assuming an incorrect IA model.

The same effect holds when our observables are extended to include shear-shear, position-position and shear-position correlations. 
The extra observables increase constraining power so that we are able to produce unbiased constraints on $\epsilon\epsilon+n\epsilon+nn$ which recover 40\% of the constraining power of the highly biased constraint from $\epsilon\epsilon$ alone when the wrong IA model is assumed.

The LA model has allowed a firm foothold on the study of IAs to develop over the last decade. We have detailed an updated implementation of the most used IA model and its physical motivation, showing a reduction in biasing of cosmological constraints. In addition we reiterate that a robust nuisance parameter model can control the biasing due to IAs for either the old or new implementations. Together, these results should give us confidence that the IA effect is under control as we begin to analyse the first data from large cosmic shear surveys.

However our somewhat pessimistic approach carries the cost of reduced constraining power. Future work, focused on accurate simulation and measurement of IAs is sure to provide more detail on the physical mechanisms responsible for the initial intrinsic ellipticity distribution and its evolution. This better knowledge of IAs will improve our ability to model them and reduce our dependence on brute-force marginalisation over nuisance parameters. As such the marginalised constraints which we present in Fig.~\ref{fig:bias_oldnew} may be a worst-case scenario. With better knowledge of IAs, better constraints on cosmology will be possible.

\section*{Acknowledgements}
The authors are very grateful to Lisa Voigt who was intimately involved as a contributor to the preparatory work leading to this paper. The authors are thankful to Benjamin Joachimi for useful discussions regarding \cite{joachimi_bridle_2009}. We also thank Rachel Bean and Istvan Laszlo for very useful discussion and work on the best implementation of the lA model. Sarah Bridle acknowledges support from the Royal Society in the form of a University Research Fellowship and the European Research Council in the form of a Starting Grant with number 240672. 

\bsp

\bibliographystyle{mn2e}
\bibliography{bibliography_DK_arxivcorrected}

\section*{Appendix: History of the LA model}

\label{sec:LA_history}

The history of galaxy intrinsic alignment studies goes back to \citet{hoyle_1949} and \citet{sciama_1955}
who discussed that galaxies may acquire angular momentum through tidal torquing, in which anisotropic shear flows distort the protogalaxy and lead to a local rotation.
The theory was further developed analytically by \citet{peebles_1969}, \citet{doroshkevich_1970}
and \citet{white_1984},
and first simulated by \citet{heavensp88}.
If galaxy orientations are correlated with their angular momenta then we may expect neighbouring galaxies to be aligned~\citep[e.g.][]{vandenboschea02}.
\citet{HRH,croftm00,catelankb01} simultaneously pointed out that this could lead to a contaminating term in weak gravitational lensing, and this was investigated further in \citet{crittendennpt01,Mackeywk02,Jing02,heymansea04,hirataea04} and \citet{bridleandking}.
See \citet{schaefer08} for a more recent review.

\citet{ciotti_dutta} described a second ansatz 
to motivate the intrinsic alignments of elliptical galaxies, based on tidal stretching of the host halo by the surrounding large-scale structure.
\citet{catelankb01} showed that the correlation function of the intrinsic alignments should be proportional to that of the tidal field and thus be significant even on large scales.
This contrasts with the correlation function expected from tidal torque theory which has a higher order dependence on the tidal field and thus decays more rapidly.

\citet{HRH,croftm00} used the shapes of dark matter halos in $N$-body 
 simulations as a proxy for elliptical galaxy shapes and estimated the possible contamination to the (then) recent cosmic shear detections. They found the contamination to be a fraction of the measured signal but \citet{HRH} commented that low redshift surveys would be strongly affected and \citet{croftm00} point out that future tomographic measurements will also be significantly affected.

\citet{hiratas04} (which we are calling HS04) 
discussed both models and used the term `linear alignment model' to describe the alignment of the galaxy halo with the tidal stretch of the local gravitational potential. 
Previous to HS04 the cosmic shear contamination was believed to come from physically close galaxies due to both galaxies forming in the same large-scale structure. HS04 identified a second term due to the intrinsic alignment of one galaxy near to a large mass which would correlate with the gravitational alignment of a distant galaxy lensed by the same mass. The first of these two was labelled `intrinsic-intrinsic' (II) correlation and the second `gravitational-intrinsic' (GI) correlation.

Following the approach of \citet{catelankb01}, HS04 assume that galaxy intrinsic ellipticity follows the linear relation
\begin{equation}
\gamma^I = -\frac{C_{1}}{4\pi G}\left(\Delta^{2}_{x}-\Delta^{2}_{y},2\Delta_{x}\Delta_{y} \right)\mathcal{S}[\Phi_{P}],
\end{equation}
where $\Phi_P$ is the Newtonian potential at the time of galaxy formation, smoothed by some filter $\mathcal{S}$ that cuts off fluctuations on galactic scales, $G$ is the Newtonian gravitational constant, $\Delta$ is a comoving derivative and $C_1$ is a normalisation constant.

After relating the primordial potential to the linear matter power spectrum, HS04 calculate the II and GI power spectra \begin{eqnarray}
P_{II}^{\textrm{HS04}}(k,\chi) &=& \left(\frac{-C_{1}\bar{\rho}(z)}{\bar{D}(z)}\right)^{2}P^{\rm lin}_{\delta\delta}(k,\chi) \label{eqn:HS04_P_II_GI1}\\
P_{GI}^{\textrm{HS04}}(k,\chi) &=& -\frac{C_{1}\bar{\rho}(z)}{\bar{D}(z)}P^{\rm lin}_{\delta\delta}(k,\chi)
\label{eqn:HS04_P_II_GI2}
\end{eqnarray}
in terms of the linear theory matter power spectrum at
comoving distance $\chi$ corresponding to redshift $z$,
 $P^{\rm lin}_{\delta\delta}(k,\chi)$. Here $\bar{\rho}(z)$ is the mean density of the Universe, $\bar{D}(z) = (1+z)D(z)$ is the rescaled growth factor
$D(z)$.

Note that HS04 actually present the power spectra for $\tilde{\gamma}_I$, the density weighted intrinsic shear (and $\delta,\tilde{\gamma}_I$ for the GI term).
This is because galaxies are not randomly positioned, but are expected to form preferentially in regions of higher matter density.
This leads to higher order terms in the IA power spectra which are usually ignored because they are about an order of magnitude smaller than the leading term in the above equations~\citep{bridleandking}.

These equations were used in several publications until the discovery of missing factors  of $a^2$ which arose in the conversion factor between the density perturbation and the primordial potential. These were corrected in an erratum~\citet{hiratas10_posterratum}, which we call HS10.
In the updated version, eqns. \ref{eqn:HS04_P_II_GI1} and \ref{eqn:HS04_P_II_GI2} become
\begin{eqnarray}
P_{II}^{\textrm{HS10}}(k,\chi) &=& \left(\frac{-C_{1}\bar{\rho}(z)}{\bar{D}(z)}a^{2}\right)^{2}P^{\rm lin}_{\delta\delta}(k,\chi)\\
P_{GI}^{\textrm{HS10}}(k,\chi) &=& -\frac{C_{1}\bar{\rho}(z)}{\bar{D}(z)}a^{2}P^{\rm lin}_{\delta\delta}(k,\chi).
\label{eqn:HS04_P_II_GI_erratum}
\end{eqnarray}

Originally HS04 related their $P_{II}$ and $P_{GI}$ to the linear matter power spectrum $P^{lin}_{\delta\delta}$. It was subsequently indicated by \citet{hirataea07} that more power at small scales may fit the data better and
 \citet{bridleandking} 
proposed that the non-linear matter power spectrum be substituted for the linear in what became known and the Non-Linear Alignment (NLA) model, which was then applied in \citet{bridleandking,schneiderb09,mandelbaumea09,joachimiea_megazlrg,kirk_bs_2010,MGpaper1} and \citet{MGpaper2}.
It is this common NLA approach, which we describe as HS04NL, 
\begin{eqnarray}
P_{II}^{\textrm{HS04NL}}(k,\chi) &=& \left(\frac{-C_{1}\bar{\rho}(z)}{\bar{D}(z)}\right)^{2}P_{\delta\delta}(k,\chi)\label{eqn:HS04_P_II_GI_NL1}\\
P_{GI}^{\textrm{HS04NL}}(k,\chi) &=& -\frac{C_{1}\bar{\rho}(z)}{\bar{D}(z)}P_{\delta\delta}(k,\chi),
\label{eqn:HS04_P_II_GI_NL2}
\end{eqnarray}
where the non-linear matter power spectrum is used in both equations. 

In this work 
we employ a more physically motivated approach based on the idea that galaxy intrinsic alignments are seeded at some early epoch of galaxy formation and do not evolve with redshift.
This was the original intent of the LA model in \citet{catelankb01} and HS04. 
This principle is apparent from the original HS04 equations by the presence of the $1/D(z)$ terms which divide out the growth.
Therefore we relate our II term directly to $P^{\rm lin}_{\delta\delta}$ but the GI term picks up some contribution from the later (linear and nonlinear) growth of large scale structure via its dependence on cosmic shear.
This is included in the formalism by the geometric mean of the linear and nonlinear matter power spectra in equation \ref{eqn:P_II_GI_new2}. This was proposed and discussed further in \citet{MGpaper1,MGpaper2}.

Note that
equations \ref{eqn:HS04_P_II_GI_NL1} and \ref{eqn:HS04_P_II_GI_NL2} 
are the same as our expressions in eqns. \ref{eqn:P_II_GI_new1} and \ref{eqn:P_II_GI_new2} with
the prefactor in brackets simplified, and there is a different treatment of the linear/non-linear matter power spectra.

An improvement on the contributions at non-linear scales was developed by \citet{schneiderb09} by developing a halo model of intrinsic alignments. We do not use it here as it does not contain full flexibility of the cosmological model.

\begin{figure}
  \begin{flushleft}
    \centering
       \includegraphics[width=3.5in,height=2in]{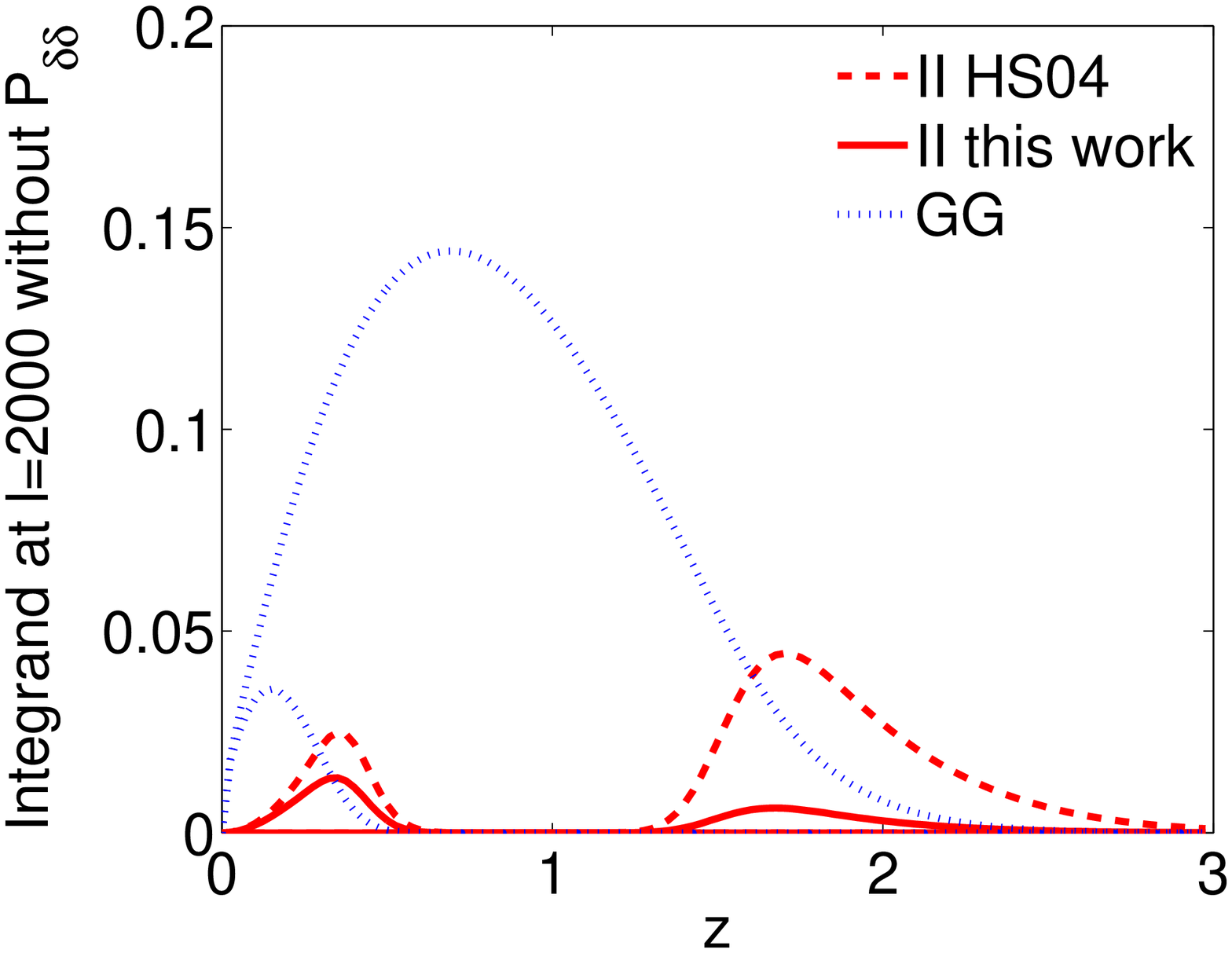}
       \includegraphics[width=3.5in,height=2in]{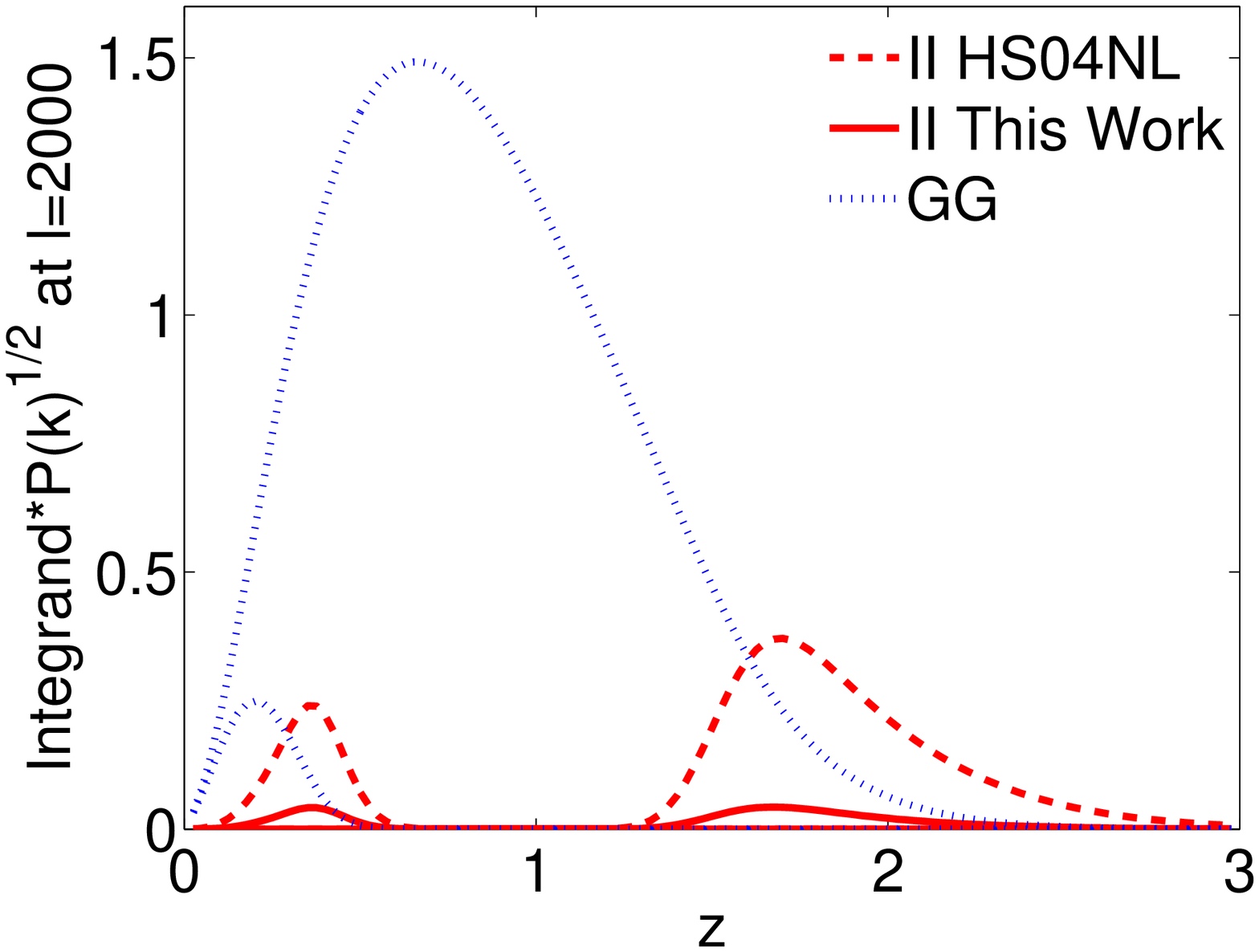}
\caption{Integrand of the GG angular power spectrum at a multipole of $l=2000$ ($2\pi G \rho(z)a^{2}c^{-2}W_{i}(\chi)\sqrt{P_{\delta\delta}(k,\chi)}$) [blue dotted] and the II integrand for the original \citet{hiratas04} LA model ($C_1 \rho(z)a n_{i}(\chi)\sqrt{P_{\delta\delta}(k,\chi)}$) [red dotted] and the LA model given in this work ($C_1 \rho(z=0) n_{i}(\chi)\sqrt{P^{lin}_{\delta\delta}(k,\chi)}$) [red solid], both shown for $l=2000$. We present integrands without [upper panel] and with the appropriate matter power spectrum term [lower panel]. Each angular power spectrum is plotted for the first and tenth tomographic redshift bins of our fiducial survey, these appear to the left and right hand of each plot respectively. Here the term $P(k)$ corresponds to either the linear or non-linear matter power spectrum depending on the term ($G$ or $I$), and on the implementation used (HS04NL or This Work).}
\label{fig:integrands}
  \end{flushleft}
\end{figure}

\begin{figure*}
  \begin{flushleft}
    \centering
       \includegraphics[width=7in,height=7in]{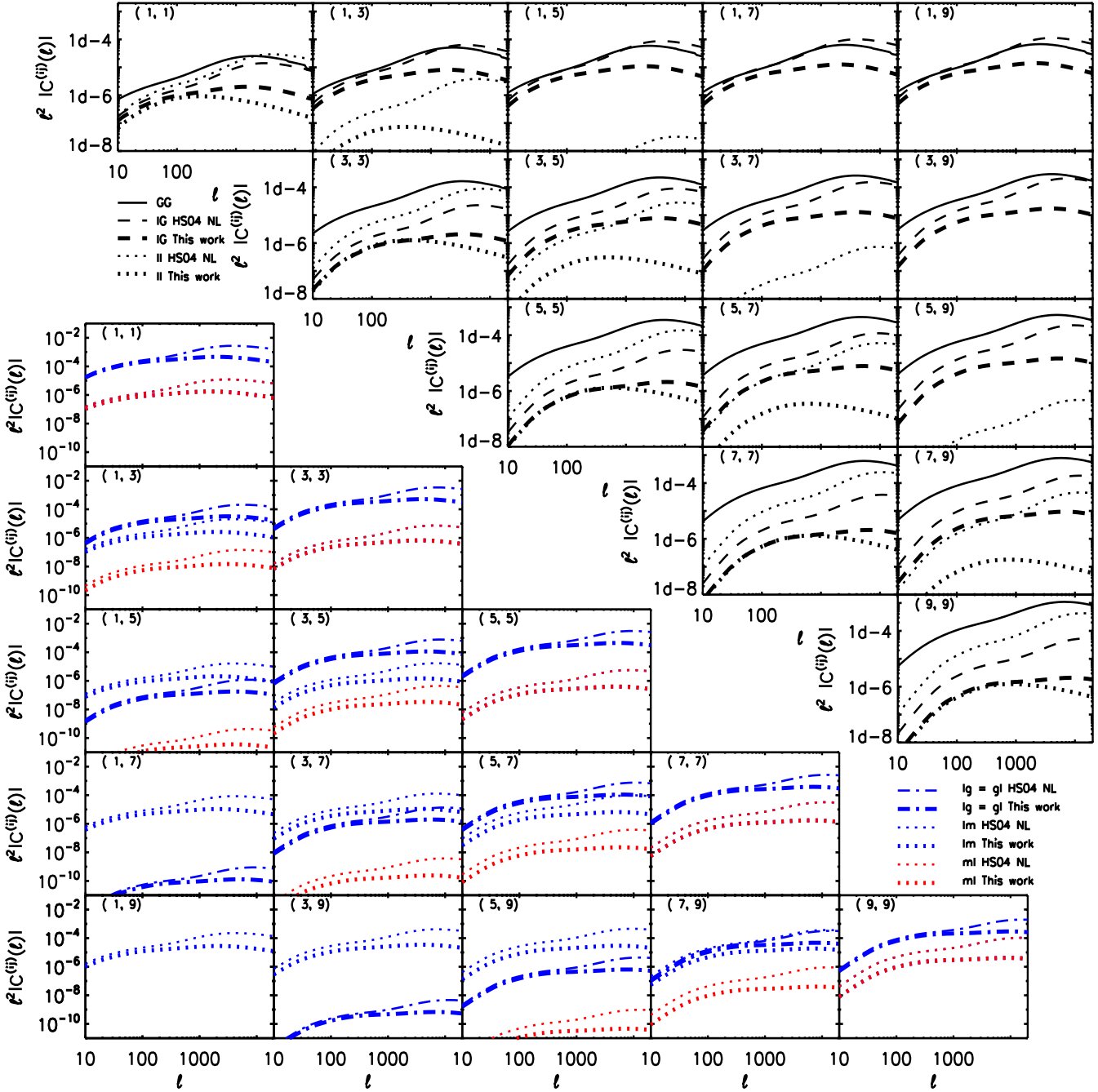}
\vspace{6mm}
\caption{Projected angular power spectra which contain IA contributions. Computed for our fiducial survey and displayed for a variety of tomographic redshift bin combinations as in Fig \ref{fig:cls_full}. Power spectra are computed using the HS04NL approach [narrow lines] and the new approach described in this work [thick lines]. The GG (solid black narrow) power spectrum is shown for comparison.}
\label{fig:cls_change}
  \end{flushleft}
\end{figure*}

In this paper we discuss and illustrate the difference between the models, 
summarised in Fig. \ref{fig:integrands}. The changes can be divided into two types:
(i) the correct use of the $a^2$ terms as included in \citet{hiratas10_posterratum} and
(ii) the consistent use of the linear matter power spectrum for the II correlation but the square root product of the linear and nonlinear power spectra for the GI correlation (see Equation \ref{eq:sqrtpk}). The upper panel of Fig. \ref{fig:integrands} includes only the first change 
by plotting the square root of the $C_l$ integrand without the contribution from the matter power spectrum. The terms plotted for bins $i=1,10$ are
\begin{eqnarray}
\textrm{GG}: &\frac{2\pi G \rho(z)a^2}{c^2}W_{i}(\chi)\\
\textrm{II}^{\textrm{HS04}}: &C_1 \rho(z)a n_{i}(\chi)\\
\textrm{II}^{\textrm{This Work}}: &C_1 \rho(z=0) n_{i}(\chi).
\end{eqnarray}
The lower panel includes the square root of the matter power spectrum contribution, plotting
\begin{eqnarray}
\textrm{GG}: &\frac{2\pi G \rho(z)a^2}{c^2}W_{i}(\chi)\sqrt{P_{\delta\delta}(k,\chi)}\\
\textrm{II}^{\textrm{HS04NL}}: &C_1 \rho(z)a n_{i}(\chi)\sqrt{P_{\delta\delta}(k,\chi)}\\
\textrm{II}^{\textrm{This Work}}: &C_1 \rho(z=0) n_{i}(\chi)\sqrt{P^{lin}_{\delta\delta}(k,\chi)}.
\end{eqnarray}

Correct inclusion of the $a^2$ terms produces an IA signal which is constant with redshift, rather than one which increases with redshift in the HS04NL prescription. As IAs are normalised against a $C_1$ value measured at low redshift this means that the new II term is comparable to the HS04NL term at low redshift (if it were possible to show the II terms at $z=0$ on the top panel of Fig. \ref{fig:integrands} they would be identical) but increasingly diverge a higher redshifts. This is why the II kernel for the 10th (high-z) redshift bin is reduced more as we move from HS04NL to the new model than the equivalent lines for the 1st (low-z) bin. As the shear-shear term is unaffected by our treatement of IAs this means that IAs make a less significant contribution to the total shear signal at higher redshift than in the HS04NL approach.

The impact of moving from the common NLA approach to our motivated treatment of nonlinear clustering and IAs is to weaken the contribution of IAs relative to cosmic shear because we discard all nonlinear clustering power for the IA terms. This means we lose all power due to nonlinear clustering in the II term and half the nonlinear power from the GI term where the I contribution is linear but the G contribution remains dependent on the fully nonlinear matter power spectrum. The impact is felt at all redshifts but the effect is strongest at low-z where nonlinear clustering is strongest. Fig. \ref{fig:cls_change} shows the projected angular power spectra for those components which are affected by the change from the HS04NL model (as used in \citet{joachimi_bridle_2009}) to the latest implementation described in this work. The format of the plot is the same as Fig. \ref{fig:cls_full}, lines corresponding to the latest LA implementation are thick versions of the same line-style as their HS04NL equivalents. The shear-shear (GG) angular power spectra are, of course, unaffected by the change, they are shown in the upper triangle for comparison. 

\label{lastpage}

\end{document}